\documentclass[12pt,preprint]{emulateapj}
\def\R25{$\rm{R}_{25}$}

\def\mum{$\mu$m }
\def\mumnospace{$\mu$m}

\def\D25{$\rm{D}_{25}$}
\def\Msun{$\rm{M}_{\odot}$}

\def\as{$^{\prime\prime}$}
\def\am{$^{\prime}$}
\def\degree{$^{\circ}$}
\def\gsim{\;\rlap{\lower 2.5pt
 \hbox{$\sim$}}\raise 1.5pt\hbox{$>$}\;}
\def\lsim{\;\rlap{\lower 2.5pt
   \hbox{$\sim$}}\raise 1.5pt\hbox{$<$}\;}

\begin{document}

\title{A pilot study using Deep Infrared Imaging to constrain the star formation history of the XUV stellar populations in NGC 4625}
\author{Stephanie J. Bush\altaffilmark{1,*}, Robert C. Kennicutt \altaffilmark{2}, M. L. N. Ashby \altaffilmark{1}, Benjamin D. Johnson \altaffilmark{2}, Fabio Bresolin \altaffilmark{3}, Giovanni Fazio\altaffilmark{1}}
\altaffiltext{1}{Harvard-Smithsonian Center for Astrophysics, 60 Garden St, Cambridge, MA 02143 USA}
\altaffiltext{2}{Institute of Astronomy, University of Cambridge, Madingley Road, Cambridge CB3 0HA UK}
\altaffiltext{3}{Institute for Astronomy, University of Hawaii, 2680 Woodlawn Drive, Honolulu, Hawaii 96822 USA}
\email{sbush@cfa.harvard.edu}
\altaffiltext{*}{Current Address: Department of Meteorology, University of Reading, Earley Gate, Reading RG6 6BB UK}
\slugcomment{Accepted for publication in ApJ}
\begin{abstract}

In a $\Lambda$CDM universe, disk galaxies' outer regions are the last to form. Characterizing their contents is critical for
understanding the ongoing process of disk formation, but observing
 outer disk stellar populations is challenging due to their low surface brightness. We
present extremely deep 3.6~\mum observations (Spitzer/IRAC) of NGC
4625, a galaxy known for its radially extended ultraviolet-emitting
stellar population. We combine the new imaging with archival UV
imaging from the {\sl GALEX} mission to derive multi-wavelength radial profiles for NGC 4625 and compare them
to stellar populations models. The colors can be explained by the
young stellar population that is responsible for the UV emission and
indicate that the current star formation rates in the outermost disk
are recent. Extended star formation in NGC 4625 may have been
initiated by an interaction with neighboring galaxies NGC 4618 and NGC
4625a, supporting
speculation that minor interactions are a common trigger for
outer disk star formation and late stage disk growth.      
\end{abstract}


\section{Introduction}

In a $\Lambda$CDM universe, the inner parts of galactic disks are
thought to assemble first. The outer parts then form gradually as
higher angular momentum material is accreted. N-body simulations of
cosmological disk formation demonstrate this ``inside out'' disk
formation, indicating that since $z\sim 1$ galaxies evolve
proportionally in scale length and mass \citep{Brook-et-al-2006,
  Mo-Mao-White-1998}.

While observations confirm stellar disk sizes grow at late
times \citep[e.g.][]{Trujillo-et-al-2006}, the mechanism for mass
buildup is unclear. Mass buildup in the outer regions of galaxies can
be accomplished several ways. Stellar mass can be redistributed
throughout a galaxy as it evolves dynamically in
isolation. \citet{Roskar-et-al-2008a} show that secular evolution,
particularly resonant scattering of stars by transient spiral arms
\citep{Sellwood-Binney-2002, Roskar-et-al-2008a}, creates sharp changes
(break radii) in the stellar profile whose radii increases with stellar age. Over time,
secular evolution populates the outer disks with older stars. Galactic winds can redistribute gas mass throughout a galaxy, preferentially driving outflows where star formation rates are high and accretion where star formation rates are low \citep{Dave-et-al-2011a, Dave-et-al-2011b}. The
accretion of low-mass satellites can dynamically induce star formation
in outer disks and deposit gaseous and stellar mass throughout the
disk \citep{Younger-et-al-2007}. Stars could also be preferentially
formed in the outer parts of galaxies at low redshift. Several
theoretical studies have shown that an increase in the specific
angular momentum of in-falling gas as a function of redshift can lead
to reservoirs of cold gas being deposited in the outer disk at late
times which  can then form stars \citep{Samland-Gerhard-2003,
  Roskar-et-al-2010, Keres-Hernquist-2009}, predicting primarily young
stellar populations in the outer regions of galaxies.

Recent imaging confirms that star formation takes place throughout
galaxies' gas disks, including the outer regions 
\citep{HerbertFort-et-al-2012, Barnes-et-al-2011}. Up to 30\% of
nearby disk galaxies have ultraviolet (UV) emission, indicating stars
$\lsim$ 200 Myrs old, in their optically faint outer regions
\citep{Thilker-et-al-2007, Zaritsky-Christlein-2007,
  Lemonias-et-al-2011}. These galaxies have been termed Extended
Ultraviolet (XUV) disks.  XUV disks are not limited to disk galaxies
with strong spiral structure. \citet{Moffett-et-al-2012} found
approximately a  40\% occurrence of XUV disks in a sample of E/S0
galaxies. Searches for star forming clusters in the outer disks of
``normal'' galaxies, galaxies not classified as XUV disks, indicate
that young stellar clusters are common to 1.3-1.5\ $R_{25}$\footnote{$R_{25}$ is
  the radius of the 25th B band magnitude isophote, and is
  traditionally regarded as the boundary between inner (stellar
  dominated) and outer (gas dominated) disks.} in disk galaxies
\citep{HerbertFort-et-al-2012,
  Barnes-et-al-2011}. \citet{MunozMateos-et-al-2007} calculated
specific star formation rate profiles (sSFR) for 161 galaxies and
found marginal evidence for a radial increase in star formation
activity, especially in more massive galaxies. However, robust
conclusions about disk formation require knowing the star formation
history of these outer regions.

The most direct way of determining the star formation history of
galaxies is to compare their old and young stellar populations. Low
levels of star formation over the lifetime of a galaxy can create a
population of old stars that has a surface brightness so low as to
make it undetectable under typical ground-based observing conditions
\citep{Bush-et-al-2008, Bush-et-al-2010}.  Despite this challenge,
observers have started to characterize outer star forming
regions in the UV, visible and infrared (IR). In M\,83, a prototypical XUV
disk \citep[\R25 = 6.9\am,][]{Thilker-et-al-2005}, the properties of
resolved Asymptotic Giant Branch (AGB) stars \citep{Davidge-2010} and
spectral energy distribution (SED) fitting of the UV and mid-IR
properties of UV-selected clumps  \citep{Dong-et-al-2008} indicate a
population of clusters with ages ranging from 200 Myrs to a few Gyrs without an
underlying older low surface brightness disk. The mid-IR properties of
UV selected clumps in the outer disks of other galaxies have yielded
similar ages \citep{Alberts-et-al-2011}. Deep B and V band imaging of
M\,101, an asymmetrical XUV disk that dominates a small galaxy cluster,
exhibits considerable azimuthal variation in its radial color profiles
with some older spiral structures and some younger structures that
were likely formed in recent interactions \citep{Mihos-et-al-2013}.

The star formation histories of outer disks have also been constrained
through measurements of their chemical abundances. Four well-known XUV
disks have been shown to have discontinuities between their inner and
outer disk abundance profiles and an approximately constant abundance
profile in the outer disk with values between approximately 0.2 and
0.33 Z$_{\odot}$ \citep{Bresolin-et-al-2009, Gildepaz-et-al-2007,
  Goddard-et-al-2010, Bresolin-et-al-2012}. While the existing gas can
be enriched to this level by constant star formation over a period of
a few Gyrs, the low stellar content in the outer disk appears
inconsistent with that level of star formation for an extended
time. This discrepancy may indicate recent gas accretion or radial gas
mixing. Surprising levels of enrichment in outer disks have been
discovered in large spirals \citep{Ferguson-et-al-1998-abundances},
gas-rich dwarfs \citep{Werk-et-al-2010-dwarfZ, Werk-et-al-2010} and
galaxies with disturbed H\,I distributions \citep{Werk-et-al-2011},
indicating this may be common.

Here we present deep {\sl Spitzer/Infrared Array Camera (IRAC)}
\citep{Fazio-et-al-2004} 3.6~\mum observations of NGC 4625. The diffuse 3.6~\mum background emission measured by Spitzer
 is a thousand times lower than
ground based K-band background emission, making it advantageous for studies of
low surface brightness stellar populations. NGC 4625 is one of two
initial prototype XUV disks with asymmetric UV emission to $4 \times
R_{25}$  
\citep[$R_{25}=47$\as,][]{Gildepaz-et-al-2005}. 
We use 3.6~\mum emission to examine how stellar light varies
throughout the disk and look for signs of a low surface
brightness infrared disk underlying the UV features. Previous studies
of outer disks in the infrared have only been sensitive to the high
surface brightness counterparts of UV clumps. We derive 3.6~\mum
profiles (Section~\ref{sec:profiles}) and compare the profile colors
to stellar populations models to constrain the star formation history
of the outer disk of the galaxy (Section~\ref{sec:pops}). We then
speculate about how these star formation histories contribute to our
understanding of galaxy formation theory.

\section{Observations and Data Reduction} \label{sec:method}

\subsection{NGC 4625}

NGC 4625 ($D=9.5$ Mpc for $H_{0} = 70$ km s$^{-1}$ Mpc$^{-1}$,
\citet{Kennicutt-et-al-2003}) is a Magellanic dwarf spiral that could be
interacting with the neighboring Magellanic dwarf NGC 4618\footnote{\citet{Odewahn1991} estimated a lower distance for NGC
  4618 of 6.0 Mpc.} or the newly discovered dwarf galaxy NGC4625A
\citep{Gildepaz-et-al-2005}. The high surface brightness inner disk is
of one-armed spiral morphology and is only a few kiloparsecs
(\R25$=1.1$\arcmin $\times 0.95$\arcmin, \cite{Kennicutt-et-al-2003})
in diameter. It has a well mapped H~I disk, coincident with the XUV
emission, of nearly constant surface density from its optical radius
to $\sim$ 5\am\, \citep{Bush-Wilcots-2004, Gildepaz-et-al-2005,
  Kaczmarek-Wilcots-2012}. A known low surface brightness optical
disk, coincident with the extended H~I, has been imaged in the B and R
bands, but without the depth needed for precise color measurements in
the outer disk \citep{Swaters-Bacells-2002, Gildepaz-et-al-2005}.  Abundances indicate a
discontinuity in metallicity at \R25 and a nearly constant outer disk
abundance of $\sim 0.2 \, Z_{\odot}$ \citep{Goddard-et-al-2011}.

\subsection{IRAC Observations} \label{sec:methodirac}

IRAC observations of NGC 4625 were carried out during Spitzer
Cycle~6 (PID\,60072, PI S.~Bush) using standard observing parameters.
Each pointing within a $2\times3$ position map having 252\as\
spacing was observed with $30\times100$\,s dithered exposures.  We
selected a cycling dither and a medium dither pattern amplitude.
The observations were configured so that NGC 4625 was centered in
the coverage map at 3.6\,$\mu$m, the more sensitive of the two operating
IRAC detectors.  Corresponding images taken at 4.5\,$\mu$m 
cover the optical disk, but not the entire field surrounding
the galaxy that we used to calculate background emission, consequently we do not use them in our analysis. NGC 4625's companion, NGC 4618, intrudes onto a portion of the images. 

The first steps in the IRAC data reduction, using the
Basic Calibrated Data (BCD), were carried out by team members
at the Center for Astrophysics. All BCD frames not containing the center of NGC 4625 were object-masked
and median-stacked according to the array detector orientation; the resulting stacked image was then visually inspected
and subtracted from the individual BCDs.  This was done to eliminate
long-term residual images arising from prior observations of bright
sources; it also served to minimize gradients in the celestial backgrounds. 
 Dust contamination can be an issue in the 3.6~\mum band \citep{Zibetti-Groves-2011},
but in the absence of corresponding images taken at longer wavelength IR images 
(unavailable in the Spitzer warm mission phase, when these observations were taken)
we cannot correct for dust contamination. Since we are 
primarily concerned with the low metallicity, low density outer regions of the galaxy, 
we would not expect this to significantly alter our results. 

After these preliminaries, the images were combined into
two spatially-registered mosaics using IRACproc \citep{Schuster-et-al-2006}.
IRACproc augments the capabilities of the standard IRAC reduction software
(MOPEX).  The software was configured to automatically flag and
reject cosmic ray hits based on pipeline-generated masks together
with a $\sigma$-clipping algorithm for spatially coincident pixels.
IRACproc calculates the spatial derivative of each image and adjusts
the clipping algorithm accordingly.  Thus, pixels where the derivative
is low (in the field) are clipped more aggressively than are pixels
where the spatial derivative is high (point sources).  This avoids
downward biasing of point source fluxes in the output mosaics.

After mosaicing, a plane was fit to the binned image to subtract a
recognizable diffuse background gradient. After masking the galaxies
in the image, the data was binned in 10 $\times$ 10 pixel bins. The
median of the resulting bins was fit with a linear fit in two
dimensions following the procedure proposed by \citet[][Appendix
  A]{Westra-et-al-2010}. The details of the background subtraction are
discussed in Appendix A.

\subsection{GALEX Observations}

NGC 4625 was imaged by Galaxy Evolution Explorer (GALEX) as a part of
the Nearby Galaxies Survey \citep{Gildepaz-et-al-2007-atlas} in the
Far-Ultraviolet (FUV) and Near Ultraviolet (NUV) bands with a total
exposure time of 3242.2 s. The exposures were processed by the GALEX
pipeline \citep[][version 6.0.1]{Martin-et-al-2005} and retrieved from
the NASA Mikulski Archive for Space Telescopes\footnote{
  http://galex.stsci.edu/GR4/?page=tilelist\&survey=allsurveys}.

Background counts in the UV images are very low, usually only a few
counts per pixel. Consequently background pixel values follow a Poisson
distribution and cannot be approximated by a symmetric, Gaussian
distribution for background subtraction. Subtracting the mean of a
Poisson distribution does not center it symmetrically around zero,
leaving a risk that the noise will bias measurements of the signal.
When the signal exceeds the background by many orders of magnitude,
the background counts can be safely ignored, but faint outer disk
features in these images are often only a few times brighter than the mean
background. Consequently, careful background subtraction is needed for
our study. We rebinned the UV images into pixels large enough that
99\% of the newly defined pixels contained at least 10 counts.  The resulting FUV pixels are 9\as\ $\times$ 9\as\
and the resulting NUV pixels are 3\as\ $\times$ 3\as (6 $\times$ 6 and 2 $\times$ 2 original pixels respectively). The
resulting pixel distribution is Gaussian and can be background
subtracted. From this point
on, only the rebinned image is used in this work.

The rebinned images were background subtracted using the same technique we
 used to subtract the background gradient  in
\S\ref{sec:methodirac}. A wide area surrounding both NGC 4625 and
its companion NGC 4618 were masked, the rebinned images were now binned for the purposes of background subtraction, and a
surface was fit to the median value of the bins following
\citet[][Appendix A]{Westra-et-al-2010}. The surface was fit
iteratively, clipping 3$\sigma$ outliers until the fit converged. The
FUV image was divided into 3 $\times$ 3 pixel (27\as\ $\times$ 27\as\ ) bins and the median of
the bins was fit with a plane. The NUV image was divided into 5
$\times$ 5 (15\as\ $\times$ 15\as\ ) pixel bins and fit with a surface of order two in both $x$
and $y$ to remove nonlinear, large scale variations in the
background. Because the galaxy was masked, there is no danger that we
subtracted the galaxy's emission with this fit.

\subsection{H\,I maps}\label{sec:hi}

H\,I maps of the NGC 4625 field were derived from observations taken
with the Very Large Array (VLA) in B and C configurations
\citep{Kaczmarek-Wilcots-2012}. The moment map was created with
natural weighting and 1800 cleaning iterations on the image file. A
flux cut off of 2$\sigma$ (0.73 \Msun/pc$^{2}$) was applied. The final
synthesized beam was  13.14\as $\times$ 13.02\as.

\section{Results} \label{sec:profiles}

\begin{figure*}
\plottwo{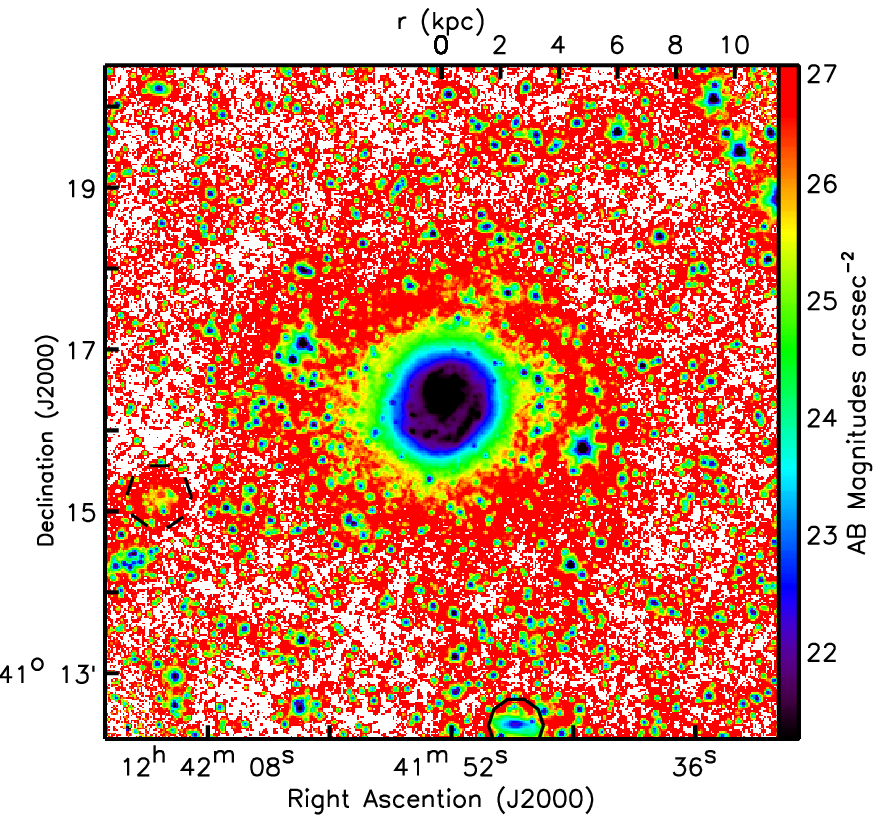}{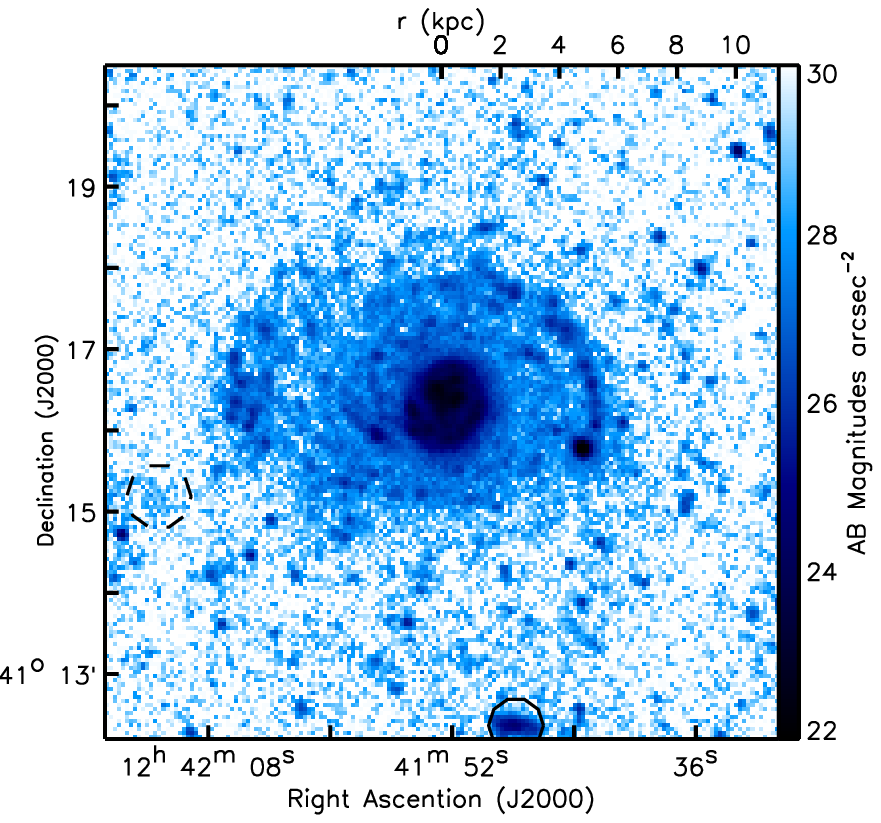}
\caption[]{Deep Spitzer/IRAC 3.6~\mum image (left) and GALEX NUV image of NGC~4625 (right). North is up and east is to the left. Two galaxies also present in the 3.6~\mum image are circled in black: NGC~4625A (dashed) and SDSS J124147.77 +411222.2 (solid). NGC~4624A is not visible in the NUV.}
\label{fig:images}
\end{figure*}
The  3.6~\mum  and NUV images of NGC~4625 are shown in
Figure~\ref{fig:images}. The filamentary NUV emission extending off
the inner disk of the galaxy is the extended ultraviolet disk. At
3.6~\mumnospace, a smooth halo of emission surrounds the
galaxy. Distant galaxies are abundant and appear as sources across the
field. These are difficult to distinguish from any possible clumpy
infrared counterparts to the NUV emission.  Also present in the image
are a few foreground stars, distinguished by the six-pointed IRAC
point response function (PRF). Finally, NGC~4625A, a low surface brightness galaxy
identified by \citet{Gildepaz-et-al-2005}, is present to the east of
the lower edge of the galaxy.

\subsection{3.6~\mum Profile} \label{sec:irprofile}

To calculate the radial profile of NGC 4625, we had to
determine which, if any, of the sources spread across the image are a
part of NGC 4625 and which are background galaxies. While in most
galaxies 3.6~\mum light is dominated by blackbody emission from older
stellar populations, stars of all ages emit at 3.6~\mum and younger AGB stars
can be brighter in the 3.6~\mum than older red giant stars. In environments possibly
dominated by younger stellar clusters, such as the outer disks, the
3.6~\mum emission could be dominated by AGB stars and it is possible
that some of the sources in the image are hot young clusters. However,
many of them must be high redshift galaxies
\citep{Fazio-et-al-2004numcounts, Ashbyetal2013}.

Due to the low resolution of the UV data (rebinned image, see \S~\ref{sec:method}) it is very difficult to
identify 3.6~\mum counterparts to UV clumps. Consequently, we could not
separate young stellar clusters from high redshift galaxies based only
on FUV, NUV and 3.6~\mum emission. Optical imaging is not available
to the depth of the 3.6~\mum data, which means we could not use the
optical colors to discriminate between background galaxies and NGC 4625 sources. It
is consequently difficult to assess the contribution these sources
have to the overall profile. Given the prevalence of background
galaxies in the 3.6~\mum \citep{Fazio-et-al-2004numcounts,
  Ashbyetal2013} and the lack of data to individually assess sources,
we chose to mask all sources in the image and
analyze only the smooth, unresolved profile of the galaxy. However, we
note that we are likely masking a few sources that are a part of NGC
4625 and consequently our derived profile is a lower limit to the
total 3.6~\mum light.

\begin{figure}
\plotone{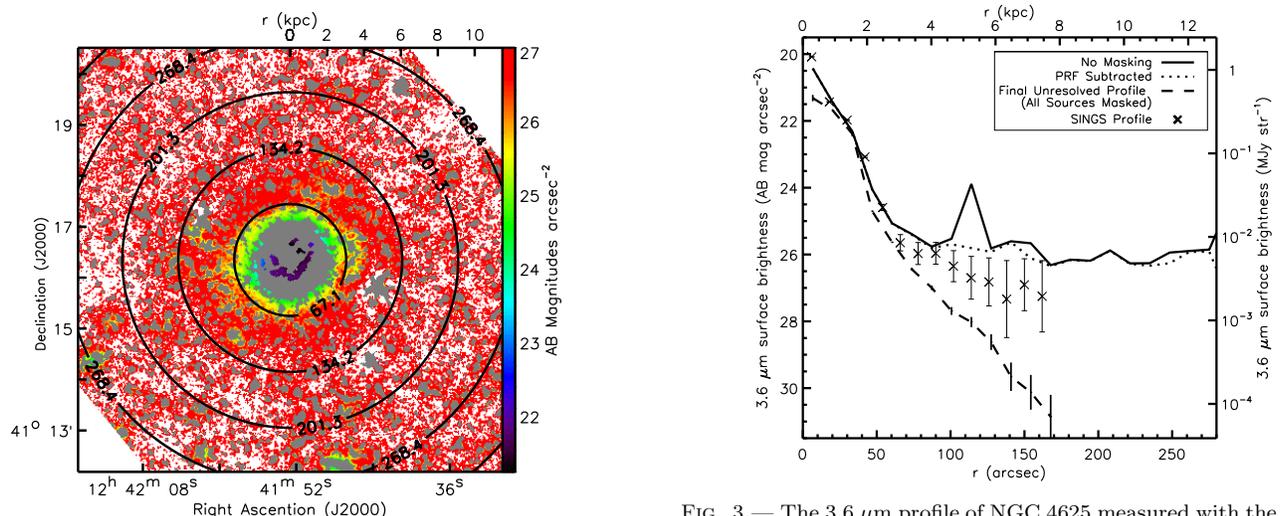}
\caption[]{Masking of sources in the 3.6~\mum image. Grey areas are masked and the color scale is identical to Figure~\ref{fig:images}. The higher surface brightness peaks in Figure~\ref{fig:images} have been masked using APEX. The remaining emission is used to measure the ``unresolved'' profile of outer disk light in Section~\ref{sec:irprofile}. Black circles mark radii with arcsecond labels. When doing aperture photometry, five apertures fit between each circle marked here.}
\label{fig:masks}
\end{figure}

We used the Astronomical Point source Extractor for MOPEX (APEX) to
detect and mask sources. APEX subtracts a local median in 15 $\times$
15 pixel windows from the image to remove any smooth variations in the
image, including the unresolved emission from the galaxy, and selects
groups of four pixels or more that are 3$\sigma$ above the local noise
value. Because the many background galaxies detected by IRAC are not
point sources, we did not attempt to fit and subtract every
source. Instead, all detected sources were masked at 3$\sigma$ above
the background noise. To be sure we were accurately accounting for the
emission in the wings of true point sources we fitted point response functions (PRFs) by eye to
sources obviously exhibiting diffraction spikes and subtracted the
fitted PRFs from the image in addition to masking their cores. The
masked image is shown in Figure~\ref{fig:masks}. This mask is
inappropriate for measuring the profile of the inner disk, because the
high surface brightness center of the galaxy is detected as a source,
so we used the profile only to analyze outer disk properties. 

\begin{figure}
\plotone{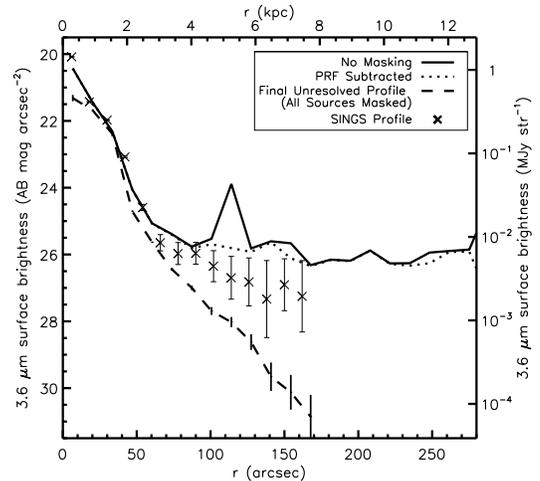}
\caption[]{The 3.6~\mum profile of NGC 4625 measured with the
  different masks or lack thereof (images and masks shown in Figures~\ref{fig:images}
  and~\ref{fig:masks}). The solid line is the profile with no
  masking. The dotted line is the profile with no masking after the
  PRFs of foreground stars have been subtracted. In these profiles,
  most of the emission beyond 70\as\, is dominated by background
  galaxies. The dashed line is the profile after all APEX detected
  sources are masked. The profiles cease where they fall to the noise
  level of the background. The 3.6~\mum profile derived from the SINGS
  survey data by \citet{MunozMateos-et-al-2009a} is shown as the crosses. For clarity, only every other point is shown. Because SINGS data
  are only $1/7$ as deep as our mosaics, the latter is much more sensitive
  to the halo of unresolved 3.6~\mum emission surrounding the galaxy,
  which is reflected in smaller error bars. Due to the depth of the
  data, we also take a different, more conservative, approach to
  masking background sources than \citet{MunozMateos-et-al-2009a},
  which we believe allows us to more accurately characterize the
  galaxy's outer disk emission and is primarily responsible for the
  difference between the two profiles (see \S~\ref{sec:irprofile}).} 
\label{fig:irprofs}
\end{figure}

Next, we measured the 3.6~\mum profile in circular annuli. Apertures
are centered on RA = 12h\,41m\,52.6s and Dec =
+41\degree\,16\am\,21.5\as\,(J2000) and are 13.5\as\, wide in order to
encompass the beam of the lowest spatial resolution data we analyze in
this study, the H~I maps (\S~\ref{sec:hi}). To evaluate the profile
error, we chose a distance beyond which we believe all the flux
is due to background sources and use apertures beyond that radius to
evaluate the error in the background. We chose the radius
corresponding to the minimum in the H\,I profile (13 kpc, derived in
Section~\ref{sec:multi}) as the edge of the galaxy and use the median
of the flux measured in annuli beyond that radius as the background
level. Error bars were derived by taking an estimate of background
uncertainty, the standard deviation of the
apertures from the median, and adding sampling errors in
quadrature. Sampling error is the error
associated with having a limited number of pixels to calculate the mean ($\sqrt{\frac{\sigma}{N}}$ where $N$ is the number of pixels and $\sigma$ is the standard deviation of their values).

The profiles are shown in Figure~\ref{fig:irprofs}. The black solid
line is the profile without any masking. Beyond 75\as, the profile
rapidly flattens because it is dominated by the contribution of
background galaxies. The dashed line is the profile with sources
masked by APEX. In the outer disk this unresolved emission falls off
sharply. 
We refer to the masked profile as the ``unresolved'' profile.

Also plotted in Figure~\ref{fig:irprofs} is the profile derived from
the SINGS survey 3.6~\mum data which is $1/7$ as deep as our data
\citep{MunozMateos-et-al-2009a}. \citet{MunozMateos-et-al-2009a}
accounted for background galaxies by applying a mask based on IRAC
colors. Without deep observations at 4.5 \mum and 8 \mum to compare to
our images, this method was not available to us, so we cannot compare
the masking methods directly. However, as their study primarily
focused on the contents of the inner disk, their masking choice was
less conservative than ours and many faint background galaxies remain
in their images after masking \citep[left side of Figure
  2]{MunozMateos-et-al-2009a}. Consequently we believe our more
conservative masking choice allows a better characterization of the
profile of the outer disk and that the difference in masking most
likely accounts for the difference between the profiles. Our
conservative masking choice is inappropriate for the inner disk, so the
\citet{MunozMateos-et-al-2009a} profile should be used to study the
inner disk. Note that at the transition between the inner and outer disk,
approximately 60\as , our profiles agree to within the error bars.

\subsection{Multi-Wavelength Profiles} \label{sec:multi}

\begin{figure}
\plotone{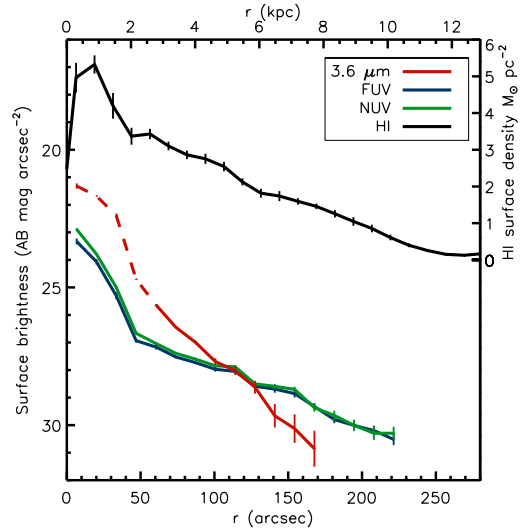}
\caption[]{Final multi-wavelength profiles after all processing
  discussed in Section~\ref{sec:multi}: H\,I (black), 3.6~\mum (red),
  FUV (blue) and NUV (green). The profiles cease when they reach the
  level of the background noise. The 3.6~\mum profile is dashed where our
  masking strategy starts to mask the inner disk's emission. All
  future figures ignore this region. The 3.6~\mum profile is only for
  the smooth component of the 3.6~\mum disk.}
\label{fig:multiprofssub}
\end{figure}

To derive colors and gas surface densities, we also calculated FUV, NUV
and H\,I profiles.
\footnote{The H\,I profile derived here disagrees with the profile previously derived from the same data by \citet{Kaczmarek-Wilcots-2012}. After a detailed discussion with the authors, they determined that the profile in \citet{Kaczmarek-Wilcots-2012} is too high by a factor of two and an erratum is in preparation. The profile derived here is correct.}
 We first transformed the UV images and H\,I map to the
same pixel scale and orientation as the 3.6~\mum image. The differing resolutions of the 3.6~\mum and UV images make it inappropriate to apply the APEX mask used for the 3.6~\mum image to the UV images. The 3.6~\mum image could be smoothed to the UV resolution prior to masking, but this makes the discrimination between 3.6~\mum sources and the 3.6~\mum smooth emission more difficult. However, the 3.6~\mum bandpass encompasses many more redshifted background galaxies than the FUV and NUV, so
background galaxies are not as prevalent in the UV. Consequently, in the UV images, we only
masked other low redshift galaxies in the field (NGC 4618, NGC 4625A)
and bright foreground stars. Then the profiles were measured with the
same apertures as in used for the 3.6~\mum profile (Section~\ref{sec:irprofile}). To subtract the
minor contribution from background galaxies, in each band we measured the median surface brightness in point-source masked annuli beyond 280\as, and subtracted the median value of these annuli from all aperture values to give the final
profiles shown in Figure~\ref{fig:multiprofssub}. The error in the
profiles is taken to be the standard deviation from this median
added in quadrature to aperture sampling errors. The
FUV and NUV profiles are consistent with the previous analysis by
\citet{Gildepaz-et-al-2005}, with an inner disk, a break, and a
shallower decline in the outer disk.
The H\,I profile shows a slow decline from the break around 50\as\, to
280\as. Because the H\,I distributions of NGC 4625 and its neighboring
galaxy, NGC 4618, overlap, the profile then begins increasing again at
280\as. For our purposes, we defined the minimum in the H\,I
profile as the edge of NGC 4625.

\begin{figure}
\plotone{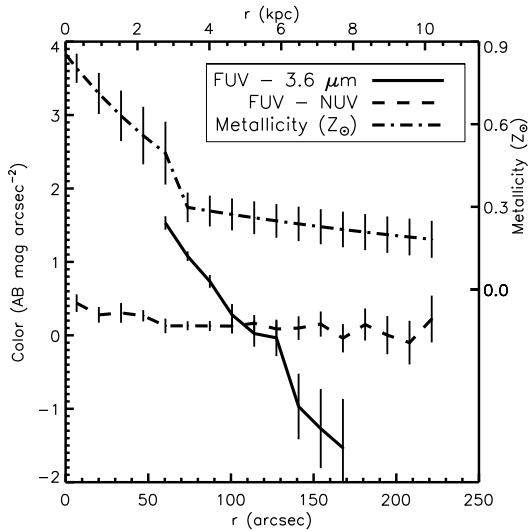}
\caption[]{FUV$-$3.6~\mum (solid) and FUV$ - $NUV (dashed) color
  profiles. They are each bluer at larger radii. The FUV$-$3.6~\mum
  colors correspond to reasonable age ranges in multiple SFH models,
  shown in Figure~\ref{fig:spops}. The metallicity profile from
  \citet{Goddard-et-al-2011} is shown as the dashed-dotted line for comparison.}
\label{fig:colorprofs}
\end{figure}

The 3.6~\mum profile is similar to the UV and H\,I profiles in that it
shows a strong break at 60\as\ and a smoothly declining outer
disk.\footnote{The 3.6~\mum profile in the inner disk is unreliable
  due to the masking, as higher surface brightness features  are
  masked. If the mask were removed from the inner disk, the break
  would be stronger. This can be verified by other published profiles
  of NGC 4625's inner disk \citep{MunozMateos-et-al-2009a}.} However,
the slope of the outer disk profile is much steeper than either the UV
or H\,I outer disk slopes. To quantitatively compare the profiles, we
show color profiles in Figure~\ref{fig:colorprofs}. While the
FUV$-$NUV color is constant with radius, the FUV$ - $3.6~\mum color
profile drops steeply from edge of the inner disk until it is no
longer detected. For comparison we plot
the gas phase metallicity gradient measured by \citet{Goddard-et-al-2011},
which, like the multi-wavelength profiles, shows a break at the edge of the inner disk.   

\section{Analysis and Discussion}

\subsection{Comparison to Stellar Populations Models} \label{sec:pops}
\begin{figure*}
\plottwo{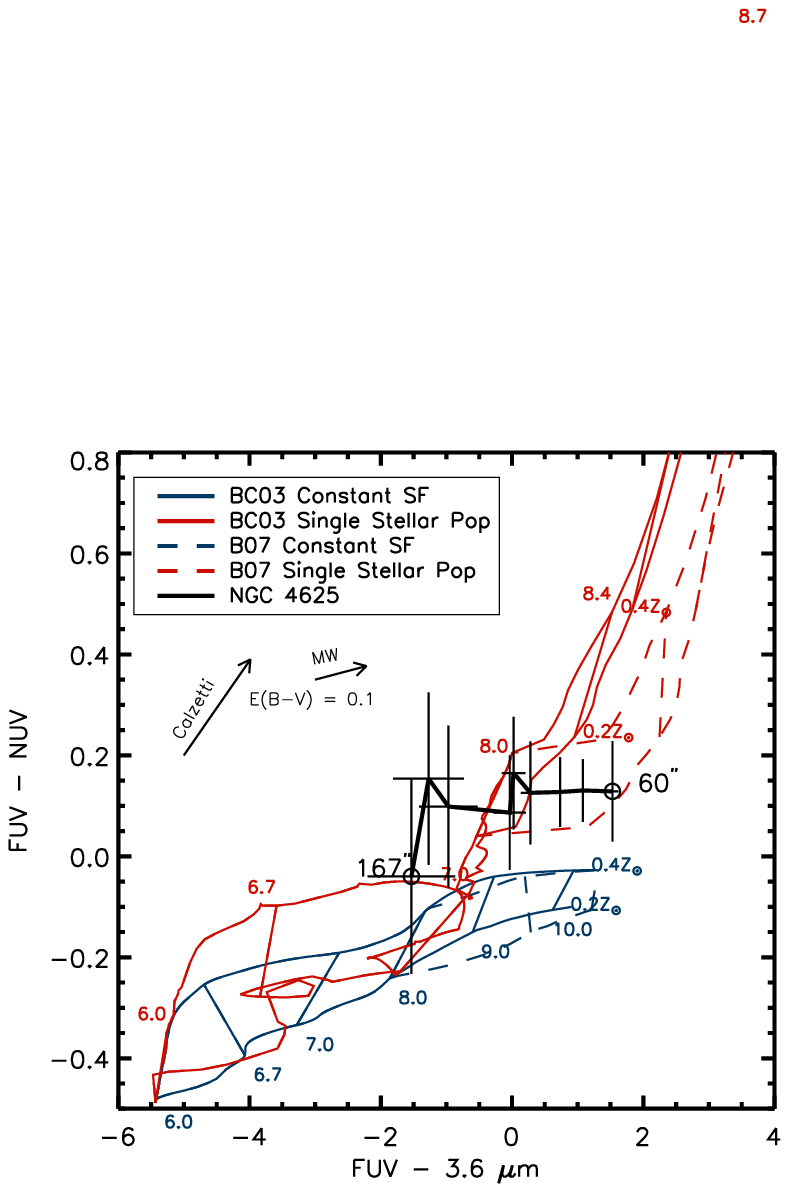}{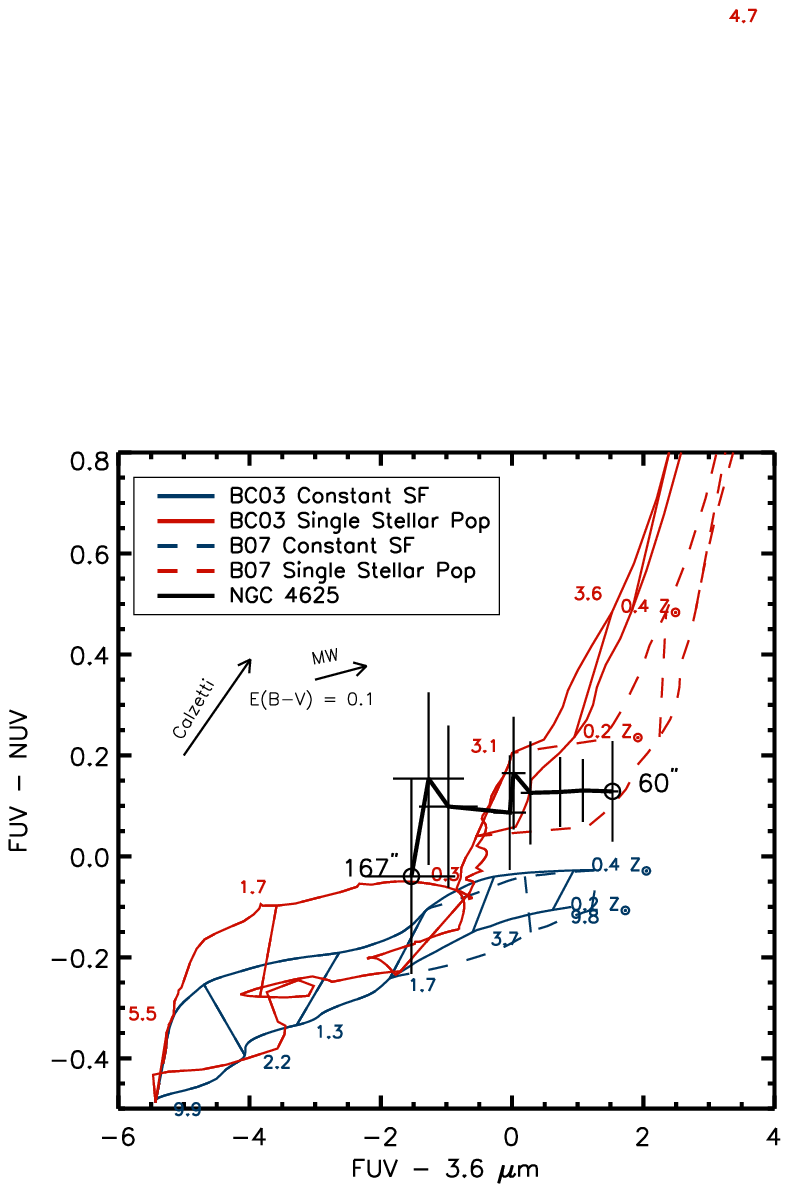}
\caption[]{The outer disk colors of NGC 4625 (black) compared to
  \citet{Bruzual-Charlot-2003} and \citet{Bruzual-Charlot-2007} stellar populations models with a constant star formation history (blue) and single stellar populations (red). On the left, numbers label log(age) in years and metallicity in solar units. On the right, numbers label 3.6~\mum mass-to-light ratio and metallicity in solar units. NGC 4625's colors are labeled with their radius in arcseconds. Extinction vectors for \citet{Calzetti-et-al-2000} and \citet{Fitzpatrick-1999} Milky Way dust extinction curves for $E(B-V) = 0.1$ are shown.  }
\label{fig:spops}
\end{figure*}
To constrain the star formation history of the outer disk of NGC 4625,
we compared the derived colors to stellar population models. These
models predict the emission of a stellar population with a given star
formation history, metallicity and initial mass function (IMF). We
chose the canonical \citet{Bruzual-Charlot-2003} models, which
we believe have the most accurate representation of the near-infrared (JHK)
contribution of AGB stars. However, we note that the contribution of
AGB stars to 3.6~\mum emission is not well understood and to estimate
the error in the models we also consider \citet{Bruzual-Charlot-2007}
models, which differ from
\citet{Bruzual-Charlot-2003} models by up to a magnitude in the 3.6~\mumnospace.

We first compared model colors for two extreme star formation rate histories,
a constant star formation rate and an instantaneous burst of star
formation (simple stellar population), to our colors. We selected a
Chabrier IMF \citep{Chabrier-2003} with mass limits of 0.1 and 100
M$_{\odot}$ and two stellar metallicities based on the gas phase
metallicities derived for NGC 4625's outer disk
\citep{Goddard-et-al-2011}. Using the [NII]/[OII] metallicity
indicator calibrated by \citet{Bresolin-2007},
\citet{Goddard-et-al-2011} measured that the outer disk abundance ranges from $0.4\,
Z_{\odot}$ to $0.2\, Z_{\odot}$. The comparison is shown in
Figure~\ref{fig:spops}, where \citet{Bruzual-Charlot-2003} models are
the solid lines and \citet{Bruzual-Charlot-2007} models are the dashed
lines. Extinction vectors for both a \citet{Calzetti-et-al-2000} and a
\citet{Fitzpatrick-1999} Milky Way dust extinction curve for $E(B-V) =
0.1$ \citep{Gildepaz-et-al-2005} are shown.

\subsection{Outer Disk Star Formation History}  \label{sec:sfh}

Figure~\ref{fig:spops} shows that the FUV $-$ 3.6~\mum color provides
the stronger constraint on the age of the stellar population and the
FUV $-$ NUV color provides the stronger constraint on the star formation history. The FUV$-$NUV
colors, particularly near the edge of the inner disk, are mostly too
red to have had a constant star formation history, but towards the
edge of the outer disk the FUV$-$NUV error bars overlap with the
constant star formation rate models. Once a star formation history is
chosen, the FUV $-$ 3.6~\mum color constrains the age. In either case,
the declining FUV$-$3.6~\mum color indicates that the age of the outer
disk decreases with radius. In a disk that has undergone a burst, the
age of the disk declines with radius from approximately 100 Myr to 10
Myr. In a disk that is constantly star forming, the age declines from
approximately 10 Gyrs to 100 Myrs. 

Galactic extinction may alter NGC 4625's colors. The
extinction vectors in Figure~\ref{fig:spops}
indicate that observed extinction curves disagree on the amount of extinction in the
FUV$-$NUV, which varies from almost none to 0.2 magnitudes, but they agree fairly well
on FUV$-$3.6~\mum extinction of 0.7 to 1.0 magnitudes. Accounting for
this extinction, the outer disk colors are more consistent with a single burst
of star formation model that has a younger age, but are also more consistent
with a constant star formation rate model. It could be argued
that applying \citet{Calzetti-et-al-2000} extinction makes the  colors of the
inner radii of the outer disk too
red to concur with a burst of star formation, but towards the inner disk the
population will undoubtedly begin to be a composite of the inner and
outer disk stellar populations, making interpretation difficult. 

Reddening within NGC 4625 could also alter the colors. Using the H$_{2}$ and H~I measurements from
\citet{Schrubaetal2011} we made a back-of-the-envelope calculation of the amount of extinction expected
using the solar neighborhood value of N(H)/E(B-V) = $5.8\times10^{21}$
\citep{Bohlinetal1978}. Depending on the reddening curve used, there
could be approximately 0.5 to 0.8 E(FUV-3.6) at 60\as\ and approximately 0.2 to 0.3 E(FUV-3.6) at 170\as. The maximum
difference between 60\as\ and 170\as\ is then 0.5 magnitudes of reddening,
which is quite small compared to the 3.5 magnitude drop in
color. Consequently, we do not believe the radial gradient in color is
due to a radial gradient in extinction.

Existing NUV$-B$ and $B-R$ color profiles
\citep{Gildepaz-et-al-2005} are not precise enough to constrain the
star formation history far into the outer disk. Near the edge of the
inner disk they agree with the UV and 3.6~\mum colors presented here: slightly too
red to be constantly star forming and consistent with single stellar
populations whose age is loosely constrained between 1 Gyr and 100
Myrs.

While the errors allow for multiple interpretations, taken at face value, the colors
of the outer disk imply a stellar population younger than 1 Gyr that
decreases in age with radius. In other words, we have not discovered a
reservoir of stellar mass that cannot be explained
by the young population that is also responsible for the UV
emission. However, we have compared 
the entire FUV profile to the smooth component of the 3.6~\mum
profile, which may not be a fair comparison. 
As discussed in \S~\ref{sec:irprofile}, without deeper optical data or higher
resolution UV data to selectively mask 3.6~\mum sources, it is very difficult to determine what portion of the
3.6~\mum sources seen near the galaxy are counterparts to young UV
clusters. 
By comparing the colors derived from the unresolved 3.6~\mum profile to stellar populations models, 
we implicitly assumed that no 3.6~\mum sources that 
belong to the galaxy were masked and that the UV and 3.6~\mum emission are from the same stellar population. 
If the 3.6~\mum profile is missing some emission due to
excessive masking, we would have underestimated the age of the
disk. 

It is also possible, even likely, that we are detecting a ``composite
population'' of a 
young outer disk stellar population
superimposed on an older, fainter population that extends from the
inner disk to the outer disk. This would cause redder colors with
decreasing radius as seen in NGC 4625's profiles. 
Instead of assuming the unresolved 3.6~\mum emission and the UV
emission are from the same stellar population, we could have assumed that we have masked all 
the counterparts to the UV emission, because they are not smooth, and
that our smooth profile reflects only a different, older
population. Of course, the truth is likely in between, but considering both allows us to determine which
conclusions are robust to our assumptions.

\begin{figure*}
\plottwo{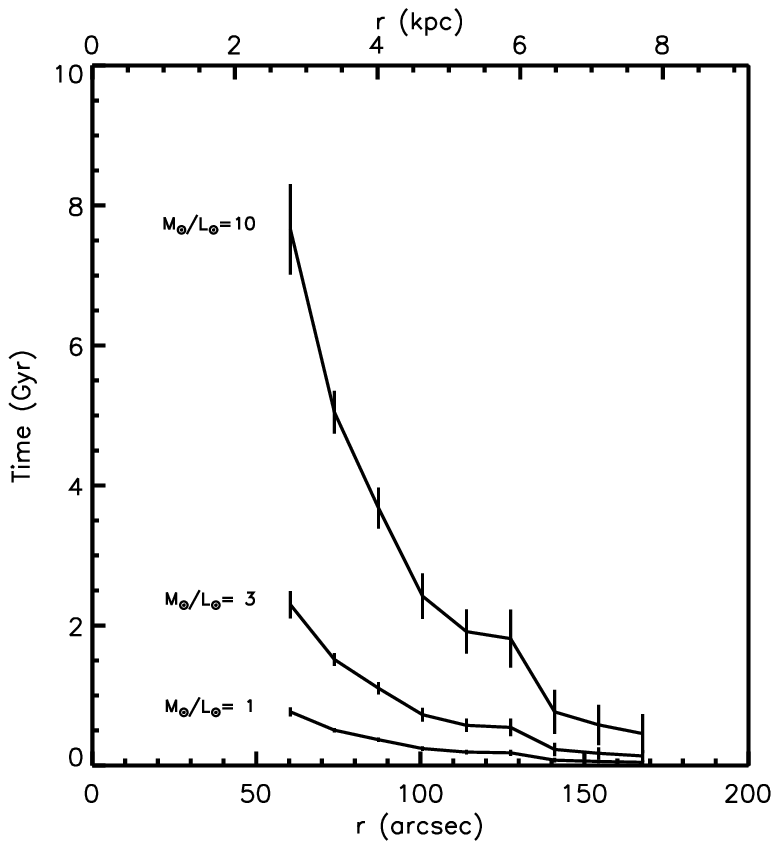}{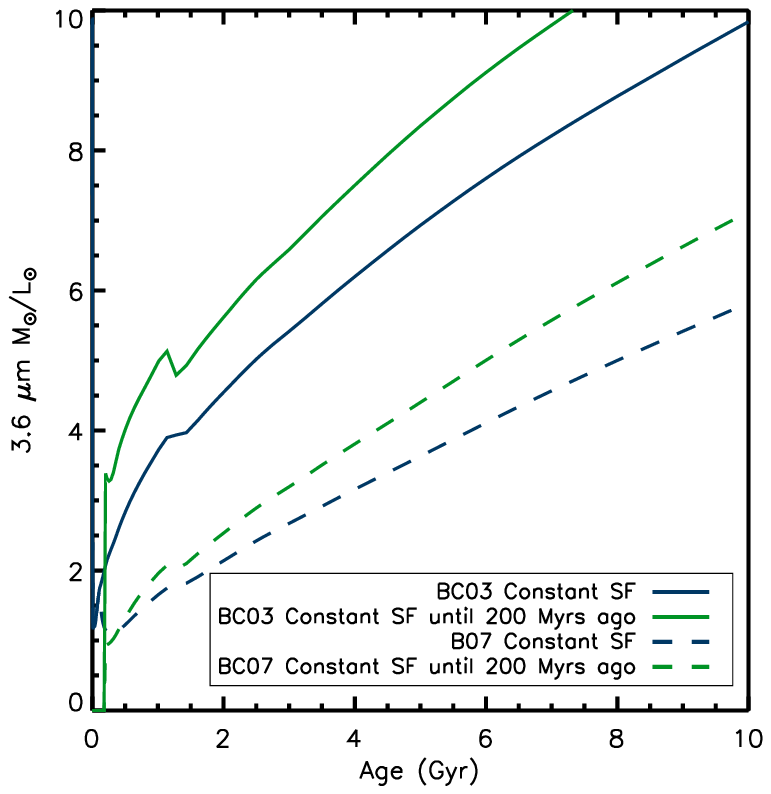}
\caption[]{ Right: The timescale to build up the mass in the outer disk using different 3.6~\mum mass-to-light ratios. Left: The 3.6~\mum mass to light ratio as a function of age with a constant star formation rate history from  \citet{Bruzual-Charlot-2003} (solid) and \citet{Bruzual-Charlot-2007} (dashed) stellar populations models. In blue the star formation rate carries on to the present day, in green all star formation ceases 200 Myr ago, approximating the non UV-emitting stellar population. }
\label{fig:timescales}
\end{figure*}

In Figure~\ref{fig:timescales} we have calculated how long the current
star formation rate, derived from the UV profile
\citep{Salim-et-al-2007}, must have been sustained in order to build the mass profile, which is derived from the
3.6~\mum emission assuming reasonable 3.6~\mum mass to light ratios
from Figure~\ref{fig:spops}. In Figure~\ref{fig:timescales}, we
have also shown how mass to light ratios evolve with age, using stellar
populations models with a constant star formation rate until the present and
stellar populations models with a constant star formation rate until
200 Myrs ago. The second model is designed to exclude the 3.6~\mum
contribution from the UV emitting population, which we have now assumed we
fully masked. At a given radius, the length of time the disk
is required to form stars at the current star formation rate to
create the smooth 3.6~\mum profile is determined by the age at which the
mass-to-light ratio and age are consistent between the two panels of Figure~\ref{fig:timescales}. At 60\as, using
\citet{Bruzual-Charlot-2003} models, these are consistent at longer timescales,
approximately 8 Gyr with higher mass to light ratios, while at 170\as,
they are consistent at less than a Gyr, with low mass to light ratios. Of
course, if the star formation rate was lower in the past, star
formation could have extended further into the past.

While the exact age of
the population is difficult to say, two conclusions seem
robust to assuming one stellar population or assuming two
independent stellar populations, one emitting in the UV and one
emitting in the 3.6~\mum. First, the mean age of the outer disk
decreases with radius. Second, the star formation at current rates in the outermost
disk is a recent event. 

While a young outer disk for NGC 4625 certainly seems plausible, there
are a few key uncertainties in this analysis. As discussed in
\S~\ref{sec:irprofile}, \citet{Zibetti-Groves-2011} have shown that
dust can make a substantial contribution to the 3.6~\mum emission,
which would violate our assumption that the 3.6~\mum emission is
emitted by stars. However, it is unlikely that there is enough hot
dust in the low density, low metallicity outer disk of NGC 4625 to
significantly alter the 3.6~\mum emission profile. The stellar populations models 
also contribute uncertainty. There is easily a factor of two difference in 
3.6~\mum mass to light ratio between \citet{Bruzual-Charlot-2003} and \citet{Bruzual-Charlot-2007}, 
and choice of the initial mass function in these models may significantly
alter the predicted emission of younger populations. 

It is well established that stars migrate radially from their birthplaces in a galaxy \citep{Roskar-et-al-2008a, Roskar-et-al-2008b} and
studies of star formation suggest that at the gas surface densities
observed in the outer disk of NGC 4625, star formation occurs with low
efficiency \citep[e.g.,][]{Martin-Kennicutt-2001}. If the stellar
content in the outer disk was not born there, our stellar populations
analysis does not reflect the characteristics of the outer disk, but
rather a scattered inner disk. However, it is unlikely that many stars
could migrate as far as 5 kpc from where they were born in less than
150 Myr. Therefore,  FUV$-$NUV colors are unaffected and suggest
that NGC 4625's young outer disk was born in-situ, in a low-efficiency
sub-threshold regime \citep{Martin-Kennicutt-2001,
  Kennicutt-1989}. However, it is possible that stellar migration from
the inner disk has created a low mass older outer disk component
coincident with the more recent star formation.   
Deep optical imaging would
further constrain whether the
outer disk is a single young population of decreasing age with radius,
or a composite population of decreasing average age, but it will not
reveal whether this population was born in-situ, or has migrated from
the inner disk.

It is difficult to determine conclusively if NGC 4625 is interacting
with either of its neighbors: NGC 4618, which is about half its mass
\citep{Bush-Wilcots-2004, Kaczmarek-Wilcots-2012} or the dwarf galaxy
NGC 4625A.  A recent interaction could have compressed gas in the
outer disk and initiated star formation in the recent past
\citep{Cox-et-al-2008, Bush-et-al-2008, Bush-et-al-2010}. The velocity
field of the outer H\,I disk is extremely regular, surprising for a
galaxy that has undergone a recent interaction. In fact, parametrized
H\,I morphology of XUV disks galaxies is rarely distinguishable from
that of other disk galaxies \citep{Holwerda-et-al-2012}. However, the star
formation rate is very low in the outer disk of NGC 4625 and  only a
small fraction of the mass needs to be compressed by tidally induced
spiral structure to create the UV emission observed
\citep{Bush-et-al-2008, Bush-et-al-2010}. If any distortion of the
disk by interaction-induced warping is present, it is undetectable in
our mosaics, because of the face-on orientation of the galaxy. Several
of the best-studied XUV disks show minor
signs of interactions, such as extended low surface brightness
features, nearby companions or warped outer H\,I disks, while having
an intact inner disk \citep{Chonis-et-al-2011, Thilker-et-al-2005,
  Mihos-et-al-2013}. This suggests the tantalizing possibility that
XUV emission is an indication of ``flyby'' encounters or interactions
with a minor perturber.  

\subsection{The Assembly of the Gaseous Outer Disk and Inside Out Disk Formation}

A young NGC 4625 outer disk is consistent with inside-out disk
formation and it seems plausible that a recent interaction with a
companion caused a burst of star formation in the existing outer gas
disk. However, it is still difficult to determine how the extremely
extended outer gas disk was assembled.
\begin{figure}
\plotone{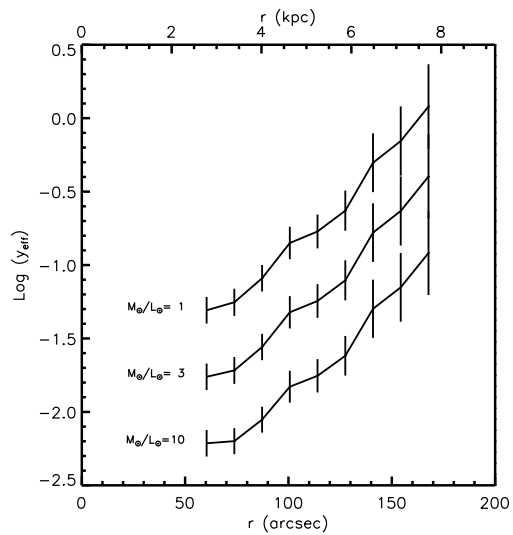}
\caption[]{The effective metal yield derived from a closed box model and assuming different 3.6
  \mum mass to light ratios. Error bars
  represent the errors in the contributing profiles. 
  }
\label{fig:yield}
\end{figure}
The metallicity of the disk gives some clues to its origin. In the
closed box chemical evolution model, which assumes a galaxy does not
exchange any gas with its environment, the relationship between a
galaxy's gas fraction and metal yield depends only on the mass of
stars that have formed in it, and is independent of star formation
rate and history. Given a metal yield $y_{eff}$, metal abundance $Z$
and gas fraction $\mu$, under the closed box model $y_{eff} =
Z/\ln(\mu)$ \citep{Edmunds-1990}. We calculate the metal yield
required by the closed box model for the gas fraction observed in the
outer disk of NGC 4625 assuming H\,I is the dominant gas mass
and a range of 3.6~\mum mass to light ratio. The effective metal yield is shown in Figure~\ref{fig:yield}
with error bars reflecting the component profiles' errors. A
reasonable metal yield would imply consistency with a closed box model
and suggest that the metal content of the gas is entirely due to the
formation and evolution of the presently observed stellar content. In
NGC 4625's outer disk, depending on the mass to light ratio chosen the metal yield increases from around 0.01 at the
edge of the inner disk to 0.3 near the edge of the outer disk. These metal
yields are very high, consistent with other measurements of outer disk
yields \citep{Werk-et-al-2011, Werk-et-al-2010-dwarfZ} and measured high metallicities
in outer disks \citep{Bresolin-et-al-2009, Bresolin-et-al-2012} and inconsistent with the closed box model. This
suggests that most of the metal content in the outer disk was not
created by in-situ star formation.

\citet{Bresolin-et-al-2012} and \citet{Werk-et-al-2011}
discuss scenarios that could enrich outer disks, including gas accretion and
radial metal mixing through either turbulent processes or galactic
outflows and re-accretion. They are unable to conclusively determine
the mechanism for enriching outer disks and comment that a
combination of processes could be at work. In the particular case of
NGC 4625, it is plausible that it has exchanged some gas in an
interaction with NGC 4618, increasing the size of the outer disk and
enriching it while simultaneously initiating star formation. However,
it is difficult to explain how a large portion of enriched gas could be
exchanged without the corresponding exchange of evolved stars. NGC 4618
also has several large holes in its inner disk HI distribution,
indicating past supernovae activity \citep{Kaczmarek-Wilcots-2012}. If
the two galaxies share a common halo, it is possible supernovae in NGC
4618 have enriched the halo's intergalactic medium. Simulations have also
shown that late accretion of halo gas can appreciably increase the gas
in an outer disk at late times \citep{Keres-Hernquist-2009}, but this
is likely to be extremely low metallicity.

\section{Conclusions and Future Directions} \label{sec:conc}

We use Spitzer/IRAC to take exceptionally deep 3.6~\mum images of NGC
4625, a prototypical extended ultraviolet disk, to constrain its star
formation history and improve our understanding of late stage disk
galaxy formation.
Since the UV is emitted by stars $\lsim 200$ Myrs old and 3.6~\mum
emission is emitted by stars of all ages, UV $-$ 3.6~\mum color
reflects the ratio between recent and past star formation rates and
constrains the star formation history of galaxies. We derive the FUV
$-$ 3.6~\mum and FUV $-$ NUV color profiles of this galaxy. We find
that color and luminosity profiles indicate either a decreasing age
with radius or that we are observing a composite of distinct
stellar populations whose mass varies differently with radius, giving
the outer disk a decreasing average age with radius. Our analysis has not revealed a reservoir of stellar mass that
cannot be explained by the young population that is also responsible for
the UV emission.  The decreasing 3.6~\mum luminosity also constrains
how long star formation at the current
rate could have continued, which is less than 2 Gyr beyond approximately 1.66 \R25.
Due to a number of uncertainties, an older, constantly star
forming disk at lower star formation rates cannot be ruled out. However, the current star
formation rates in the outermost disk are recent.

A young outer disk is consistent with the tenet of $\Lambda$CDM that outer disk
formation happens at late times. Given the presence of companions, it
is plausible that star formation was initiated by an interaction a short
time ago. The outer disk is metal rich for its stellar content,
indicating that gas has been accreted from its environment or could
have been exchanged with the companion when star formation was
initiated. However, under that scenario, it is not clear why the
stars that previously enriched the accreted gas were not also accreted.

Our analysis demonstrates the potential of deep 3.6~\mum imaging of
outer disks, but also reveals several key uncertainties. Most importantly, it is difficult to clearly separate the
emission from high redshift galaxies along the line of sight from the
emission of NGC 4625 itself. Without comparably deep and high
resolution optical or UV data, such as
\citet{Mihos-et-al-2013} assembled for M101, we cannot be certain that our 3.6~\mum
profile is more than a lower limit on the emission. Additionally, the
contribution of dust and AGB star emission to the 3.6~\mum is not well
known, complicating the interpretation of existing stellar populations
models. Improved theoretical understanding and extremely deep,
multi-wavelength datasets of nearby star forming
galaxies are required to calculate precise ages of galaxies' farthest
reaches. However, these studies are crucial for understanding the
evolution of galaxy outskirts and late stage galaxy formation.

\section*{Acknowledgments}

SJB thanks the Institute of Astronomy at the University of Cambridge
for welcoming her as a visitor for a year. Special thanks go to Eduard
Westra for assistance with background subtraction, Eric Wilcots and
Jane Kaczmarek for
sharing their H\,I moment maps and Amanda Kepley
for advice on analyzing the H\,I data. SJB thanks Lars Hernquist and
David Thilker for inspiration and countless useful conversations. 
We also thank an anonymous referee for useful feedback that improved
the manuscript.
This work is based in part on observations made with the Spitzer Space
Telescope, which is operated by the Jet Propulsion Laboratory,
California Institute of Technology under a contract with NASA. 

\section*{Appendix A: Background subtracting the 3.6~\mum image}

\begin{figure*}
\plotone{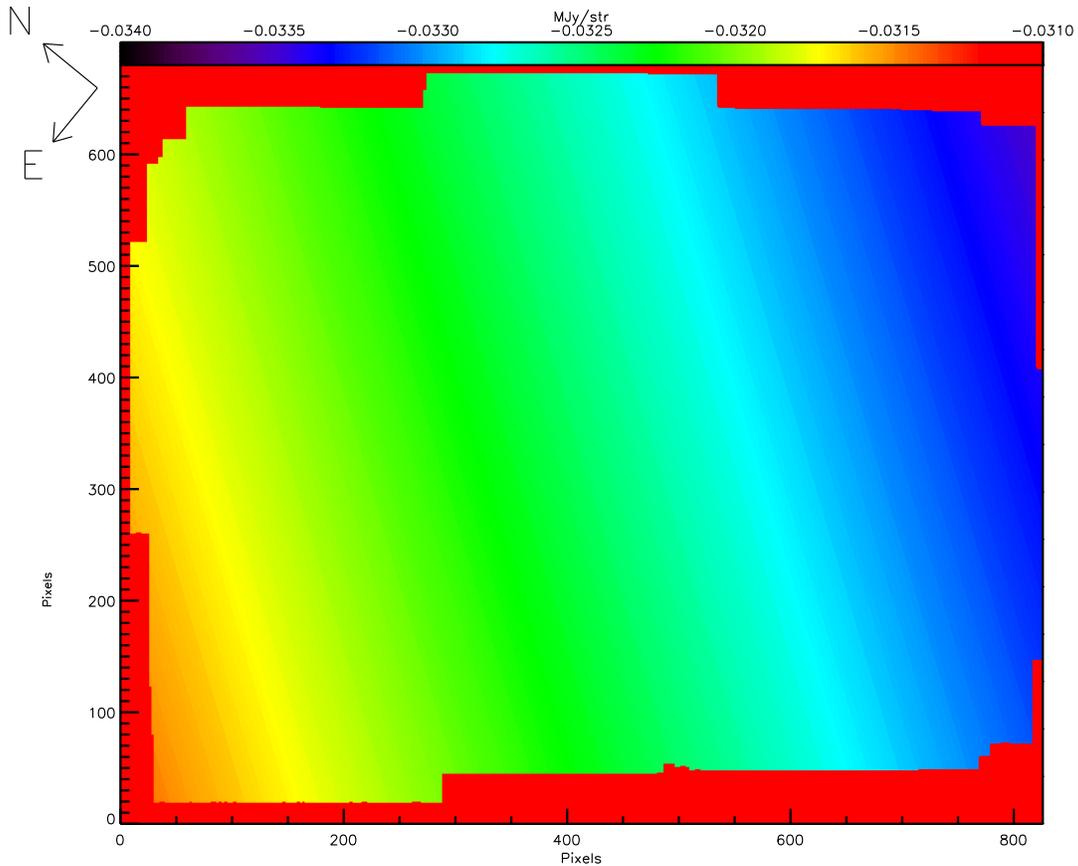}
\caption[]{The background plane subtracted from the 3.6~\mum image. The axes are in units of pixels, and one pixel equals 1.22\as\ or 56 parsecs.}
\label{fig:bkgndplot}
\end{figure*}

\begin{figure*}
\plotone{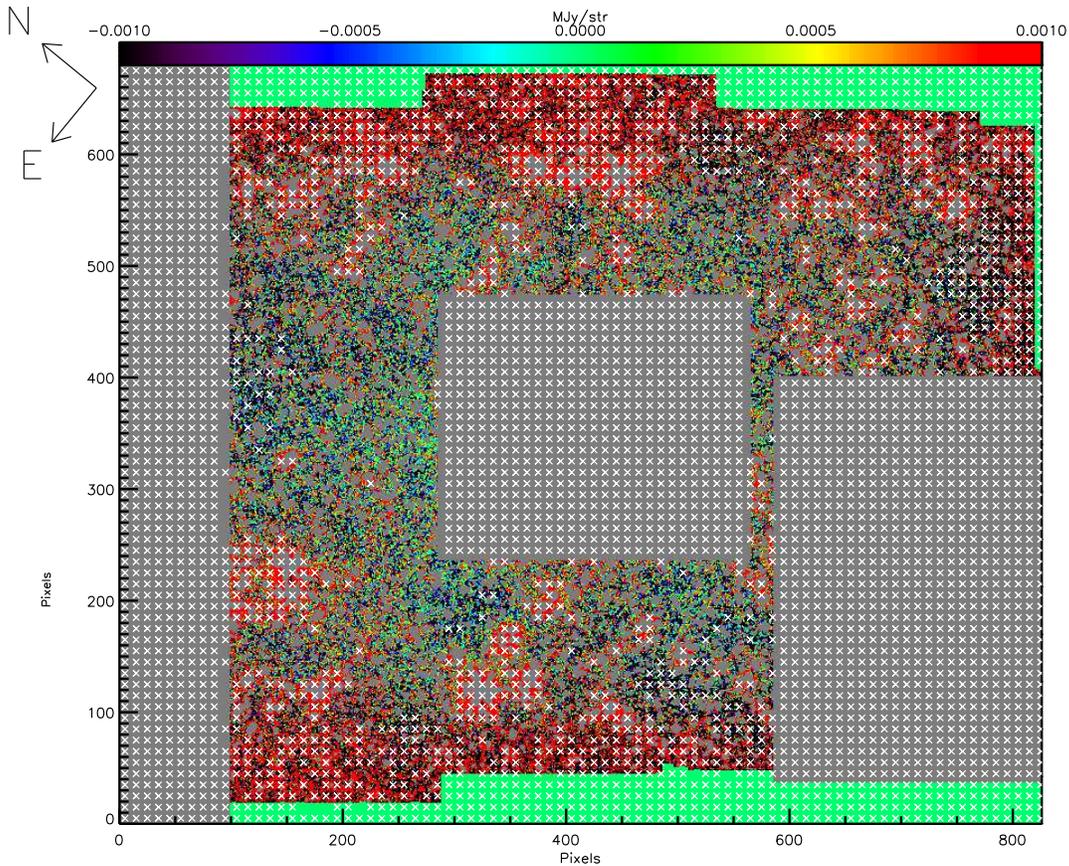}
\caption[]{The background subtracted 3.6~\mum image. The gray areas are sources that were masked either by hand or using APEX. White crosses show locations that were rejected as outliers from the fit, primarily in masked regions, at the edges of the frame where the exposure time is low due to the dither or around large point sources. Remaining colored points were fit with a plane, and the colors correspond to the post-subtraction emission. The axes are in units of pixels, and one pixel equals 1.22\as\ or 56 parsecs.}
\label{fig:bkgndfit}
\end{figure*}

Due to the sheer number of sources in this deep IRAC image and the low
surface brightness signal we are examining, background subtraction was
undertaken very carefully. First, anything that was possibly a source
was masked. NGC 4625, its companion NGC 4618 and particularly bright
point sources were generously masked by hand, and remaining sources
were masked at the 3$\sigma$ level using APEX. Then, the image was
binned in $10 \times 10$ bins, and a plane was fit to the 3$\sigma$
clipped median value of each bin. The binned data was 2$\sigma$
clipped until the fit converged. Figure~\ref{fig:bkgndplot} shows the
background subtracted from the image and Figure~\ref{fig:bkgndfit}
shows the background subtracted image, with masked areas shown in gray
and ``X''s over the rejected bins. The two large gray squares mask the
galaxies, and the large gray column to the left masks a strip with
several bright point sources. Notice that the remaining pixels show
small scatter around a value of 0 MJy str$^{-1}$. 

\bibliography{astrorefsV2}

\begin{thebibliography}{65}
\expandafter\ifx\csname natexlab\endcsname\relax\def\natexlab#1{#1}\fi

\bibitem[{{Alberts} {et~al.}(2011){Alberts}, {Calzetti}, {Dong}, {Johnson},
  {Dale}, {Bianchi}, {Chandar}, {Kennicutt}, {Meurer}, {Regan}, \&
  {Thilker}}]{Alberts-et-al-2011}
{Alberts}, S., {Calzetti}, D., {Dong}, H., {Johnson}, L.~C., {Dale}, D.~A.,
  {Bianchi}, L., {Chandar}, R., {Kennicutt}, R.~C., {Meurer}, G.~R., {Regan},
  M., \& {Thilker}, D. 2011, \apj, 731, 28

\bibitem[{{Ashby} {et~al.}(2013){Ashby}, {Willner}, {Fazio}, {Huang}, {Arendt},
  {Barmby}, {Barro}, {Bell}, {Bouwens}, {Cattaneo}, {Croton}, {Dav{\'e}},
  {Dunlop}, {Egami}, {Faber}, {Finlator}, {Grogin}, {Guhathakurta},
  {Hernquist}, {Hora}, {Illingworth}, {Kashlinsky}, {Koekemoer}, {Koo},
  {Labb{\'e}}, {Li}, {Lin}, {Moseley}, {Nandra}, {Newman}, {Noeske}, {Ouchi},
  {Peth}, {Rigopoulou}, {Robertson}, {Sarajedini}, {Simard}, {Smith}, {Wang},
  {Wechsler}, {Weiner}, {Wilson}, {Wuyts}, {Yamada}, \& {Yan}}]{Ashbyetal2013}
{Ashby}, M.~L.~N., {Willner}, S.~P., {Fazio}, G.~G., {Huang}, J.-S., {Arendt},
  R., {Barmby}, P., {Barro}, G., {Bell}, E.~F., {Bouwens}, R., {Cattaneo}, A.,
  {Croton}, D., {Dav{\'e}}, R., {Dunlop}, J.~S., {Egami}, E., {Faber}, S.,
  {Finlator}, K., {Grogin}, N.~A., {Guhathakurta}, P., {Hernquist}, L., {Hora},
  J.~L., {Illingworth}, G., {Kashlinsky}, A., {Koekemoer}, A.~M., {Koo}, D.~C.,
  {Labb{\'e}}, I., {Li}, Y., {Lin}, L., {Moseley}, H., {Nandra}, K., {Newman},
  J., {Noeske}, K., {Ouchi}, M., {Peth}, M., {Rigopoulou}, D., {Robertson}, B.,
  {Sarajedini}, V., {Simard}, L., {Smith}, H.~A., {Wang}, Z., {Wechsler}, R.,
  {Weiner}, B., {Wilson}, G., {Wuyts}, S., {Yamada}, T., \& {Yan}, H. 2013,
  \apj, 769, 80

\bibitem[{{Barnes} {et~al.}(2011){Barnes}, {van Zee}, \&
  {Skillman}}]{Barnes-et-al-2011}
{Barnes}, K.~L., {van Zee}, L., \& {Skillman}, E.~D. 2011, \apj, 743, 137

\bibitem[{{Bohlin} {et~al.}(1978){Bohlin}, {Savage}, \&
  {Drake}}]{Bohlinetal1978}
{Bohlin}, R.~C., {Savage}, B.~D., \& {Drake}, J.~F. 1978, \apj, 224, 132

\bibitem[{{Bresolin}(2007)}]{Bresolin-2007}
{Bresolin}, F. 2007, \apj, 656, 186

\bibitem[{{Bresolin} {et~al.}(2012){Bresolin}, {Kennicutt}, \&
  {Ryan-Weber}}]{Bresolin-et-al-2012}
{Bresolin}, F., {Kennicutt}, R.~C., \& {Ryan-Weber}, E. 2012, \apj, 750, 122

\bibitem[{{Bresolin} {et~al.}(2009){Bresolin}, {Ryan-Weber}, {Kennicutt}, \&
  {Goddard}}]{Bresolin-et-al-2009}
{Bresolin}, F., {Ryan-Weber}, E., {Kennicutt}, R.~C., \& {Goddard}, Q. 2009,
  \apj, 695, 580

\bibitem[{{Brook} {et~al.}(2006){Brook}, {Kawata}, {Martel}, {Gibson}, \&
  {Bailin}}]{Brook-et-al-2006}
{Brook}, C.~B., {Kawata}, D., {Martel}, H., {Gibson}, B.~K., \& {Bailin}, J.
  2006, \apj, 639, 126

\bibitem[{{Bruzual}(2007)}]{Bruzual-Charlot-2007}
{Bruzual}, G. 2007, in Astronomical Society of the Pacific Conference Series,
  Vol. 374, From Stars to Galaxies: Building the Pieces to Build Up the
  Universe, ed. {A.~Vallenari, R.~Tantalo, L.~Portinari, \& A.~Moretti}, 303--+

\bibitem[{{Bruzual} \& {Charlot}(2003)}]{Bruzual-Charlot-2003}
{Bruzual}, G. \& {Charlot}, S. 2003, \mnras, 344, 1000

\bibitem[{{Bush} {et~al.}(2010){Bush}, {Cox}, {Hayward}, {Thilker},
  {Hernquist}, \& {Besla}}]{Bush-et-al-2010}
{Bush}, S.~J., {Cox}, T.~J., {Hayward}, C.~C., {Thilker}, D., {Hernquist}, L.,
  \& {Besla}, G. 2010, \apj, 713, 780

\bibitem[{{Bush} {et~al.}(2008){Bush}, {Cox}, {Hernquist}, {Thilker}, \&
  {Younger}}]{Bush-et-al-2008}
{Bush}, S.~J., {Cox}, T.~J., {Hernquist}, L., {Thilker}, D., \& {Younger},
  J.~D. 2008, \apjl, 683, L13

\bibitem[{{Bush} \& {Wilcots}(2004)}]{Bush-Wilcots-2004}
{Bush}, S.~J. \& {Wilcots}, E.~M. 2004, \aj, 128, 2789

\bibitem[{{Calzetti} {et~al.}(2000){Calzetti}, {Armus}, {Bohlin}, {Kinney},
  {Koornneef}, \& {Storchi-Bergmann}}]{Calzetti-et-al-2000}
{Calzetti}, D., {Armus}, L., {Bohlin}, R.~C., {Kinney}, A.~L., {Koornneef}, J.,
  \& {Storchi-Bergmann}, T. 2000, \apj, 533, 682

\bibitem[{{Chabrier}(2003)}]{Chabrier-2003}
{Chabrier}, G. 2003, \pasp, 115, 763

\bibitem[{{Chonis} {et~al.}(2011){Chonis}, {Mart{\'{\i}}nez-Delgado}, {Gabany},
  {Majewski}, {Hill}, {Gralak}, \& {Trujillo}}]{Chonis-et-al-2011}
{Chonis}, T.~S., {Mart{\'{\i}}nez-Delgado}, D., {Gabany}, R.~J., {Majewski},
  S.~R., {Hill}, G.~J., {Gralak}, R., \& {Trujillo}, I. 2011, \aj, 142, 166

\bibitem[{{Cox} {et~al.}(2008){Cox}, {Jonsson}, {Somerville}, {Primack}, \&
  {Dekel}}]{Cox-et-al-2008}
{Cox}, T.~J., {Jonsson}, P., {Somerville}, R.~S., {Primack}, J.~R., \& {Dekel},
  A. 2008, \mnras, 384, 386

\bibitem[{{Dav{\'e}} {et~al.}(2011{\natexlab{a}}){Dav{\'e}}, {Finlator}, \&
  {Oppenheimer}}]{Dave-et-al-2011b}
{Dav{\'e}}, R., {Finlator}, K., \& {Oppenheimer}, B.~D. 2011{\natexlab{a}},
  \mnras, 416, 1354

\bibitem[{{Dav{\'e}} {et~al.}(2011{\natexlab{b}}){Dav{\'e}}, {Oppenheimer}, \&
  {Finlator}}]{Dave-et-al-2011a}
{Dav{\'e}}, R., {Oppenheimer}, B.~D., \& {Finlator}, K. 2011{\natexlab{b}},
  \mnras, 415, 11

\bibitem[{{Davidge}(2010)}]{Davidge-2010}
{Davidge}, T.~J. 2010, \apj, 718, 1428

\bibitem[{{Dong} {et~al.}(2008){Dong}, {Calzetti}, {Regan}, {Thilker},
  {Bianchi}, {Meurer}, \& {Walter}}]{Dong-et-al-2008}
{Dong}, H., {Calzetti}, D., {Regan}, M., {Thilker}, D., {Bianchi}, L.,
  {Meurer}, G.~R., \& {Walter}, F. 2008, \aj, 136, 479

\bibitem[{{Edmunds}(1990)}]{Edmunds-1990}
{Edmunds}, M.~G. 1990, \mnras, 246, 678

\bibitem[{{Fazio} {et~al.}(2004{\natexlab{a}})}]{Fazio-et-al-2004numcounts}
{Fazio}, G.~G. {et~al.} 2004{\natexlab{a}}, \apjs, 154, 39

\bibitem[{{Fazio} {et~al.}(2004{\natexlab{b}})}]{Fazio-et-al-2004}
---. 2004{\natexlab{b}}, \apjs, 154, 10

\bibitem[{{Ferguson} {et~al.}(1998){Ferguson}, {Gallagher}, \&
  {Wyse}}]{Ferguson-et-al-1998-abundances}
{Ferguson}, A.~M.~N., {Gallagher}, J.~S., \& {Wyse}, R.~F.~G. 1998, \aj, 116,
  673

\bibitem[{{Fitzpatrick}(1999)}]{Fitzpatrick-1999}
{Fitzpatrick}, E.~L. 1999, \pasp, 111, 63

\bibitem[{{Gil de Paz} {et~al.}(2005)}]{Gildepaz-et-al-2005}
{Gil de Paz}, A. {et~al.} 2005, \apjl, 627, L29

\bibitem[{{Gil de Paz} {et~al.}(2007{\natexlab{a}})}]{Gildepaz-et-al-2007}
---. 2007{\natexlab{a}}, \apj, 661, 115

\bibitem[{{Gil de Paz}
  {et~al.}(2007{\natexlab{b}})}]{Gildepaz-et-al-2007-atlas}
---. 2007{\natexlab{b}}, \apjs, 173, 185

\bibitem[{{Goddard} {et~al.}(2011){Goddard}, {Bresolin}, {Kennicutt},
  {Ryan-Weber}, \& {Rosales-Ortega}}]{Goddard-et-al-2011}
{Goddard}, Q.~E., {Bresolin}, F., {Kennicutt}, R.~C., {Ryan-Weber}, E.~V., \&
  {Rosales-Ortega}, F.~F. 2011, \mnras, 412, 1246

\bibitem[{{Goddard} {et~al.}(2010){Goddard}, {Kennicutt}, \&
  {Ryan-Weber}}]{Goddard-et-al-2010}
{Goddard}, Q.~E., {Kennicutt}, R.~C., \& {Ryan-Weber}, E.~V. 2010, \mnras, 405,
  2791

\bibitem[{{Herbert-Fort} {et~al.}(2012){Herbert-Fort}, {Zaritsky}, {Moustakas},
  {Di Paola}, {Pogge}, \& {Ragazzoni}}]{HerbertFort-et-al-2012}
{Herbert-Fort}, S., {Zaritsky}, D., {Moustakas}, J., {Di Paola}, A., {Pogge},
  R.~W., \& {Ragazzoni}, R. 2012, \apj, 754, 110

\bibitem[{{Holwerda} {et~al.}(2012){Holwerda}, {Pirzkal}, \&
  {Heiner}}]{Holwerda-et-al-2012}
{Holwerda}, B.~W., {Pirzkal}, N., \& {Heiner}, J.~S. 2012, \mnras, 427, 3159

\bibitem[{{Kaczmarek} \& {Wilcots}(2012)}]{Kaczmarek-Wilcots-2012}
{Kaczmarek}, J.~F. \& {Wilcots}, E.~M. 2012, \aj, 144, 67

\bibitem[{{Kennicutt}(1989)}]{Kennicutt-1989}
{Kennicutt}, Jr., R.~C. 1989, \apj, 344, 685

\bibitem[{{Kennicutt} {et~al.}(2003)}]{Kennicutt-et-al-2003}
{Kennicutt}, Jr., R.~C. {et~al.} 2003, PASP, 115, 928

\bibitem[{{Kere{\v s}} \& {Hernquist}(2009)}]{Keres-Hernquist-2009}
{Kere{\v s}}, D. \& {Hernquist}, L. 2009, \apjl, 700, L1

\bibitem[{{Lemonias} {et~al.}(2011){Lemonias}, {Schiminovich}, {Thilker},
  {Wyder}, {Martin}, {Seibert}, {Treyer}, {Bianchi}, {Heckman}, {Madore}, \&
  {Rich}}]{Lemonias-et-al-2011}
{Lemonias}, J.~J., {Schiminovich}, D., {Thilker}, D., {Wyder}, T.~K., {Martin},
  D.~C., {Seibert}, M., {Treyer}, M.~A., {Bianchi}, L., {Heckman}, T.~M.,
  {Madore}, B.~F., \& {Rich}, R.~M. 2011, \apj, 733, 74

\bibitem[{{Martin} \& {Kennicutt}(2001)}]{Martin-Kennicutt-2001}
{Martin}, C.~L. \& {Kennicutt}, Jr., R.~C. 2001, \apj, 555, 301

\bibitem[{{Martin} {et~al.}(2005)}]{Martin-et-al-2005}
{Martin}, D.~C. {et~al.} 2005, \apjl, 619, L1

\bibitem[{{Mihos} {et~al.}(2013){Mihos}, {Harding}, {Spengler}, {Rudick}, \&
  {Feldmeier}}]{Mihos-et-al-2013}
{Mihos}, C., {Harding}, P., {Spengler}, C., {Rudick}, C., \& {Feldmeier}, J.
  2013, \apj, 762, 1

\bibitem[{{Mo} {et~al.}(1998){Mo}, {Mao}, \& {White}}]{Mo-Mao-White-1998}
{Mo}, H.~J., {Mao}, S., \& {White}, S.~D.~M. 1998, \mnras, 295, 319

\bibitem[{{Moffett} {et~al.}(2012){Moffett}, {Kannappan}, {Baker}, \&
  {Laine}}]{Moffett-et-al-2012}
{Moffett}, A.~J., {Kannappan}, S.~J., {Baker}, A.~J., \& {Laine}, S. 2012,
  \apj, 745, 34

\bibitem[{{Mu{\~n}oz-Mateos} {et~al.}(2007){Mu{\~n}oz-Mateos}, {Gil de Paz},
  {Boissier}, {Zamorano}, {Jarrett}, {Gallego}, \&
  {Madore}}]{MunozMateos-et-al-2007}
{Mu{\~n}oz-Mateos}, J.~C., {Gil de Paz}, A., {Boissier}, S., {Zamorano}, J.,
  {Jarrett}, T., {Gallego}, J., \& {Madore}, B.~F. 2007, \apj, 658, 1006

\bibitem[{{Mu{\~n}oz-Mateos} {et~al.}(2009)}]{MunozMateos-et-al-2009a}
{Mu{\~n}oz-Mateos}, J.~C. {et~al.} 2009, \apj, 703, 1569

\bibitem[{{Odewahn}(1991)}]{Odewahn1991}
{Odewahn}, S.~C. 1991, \aj, 101, 829

\bibitem[{{Ro{\v s}kar} {et~al.}(2010){Ro{\v s}kar}, {Debattista}, {Brooks},
  {Quinn}, {Brook}, {Governato}, {Dalcanton}, \& {Wadsley}}]{Roskar-et-al-2010}
{Ro{\v s}kar}, R., {Debattista}, V.~P., {Brooks}, A.~M., {Quinn}, T.~R.,
  {Brook}, C.~B., {Governato}, F., {Dalcanton}, J.~J., \& {Wadsley}, J. 2010,
  \mnras, 408, 783

\bibitem[{{Ro{\v s}kar} {et~al.}(2008{\natexlab{a}}){Ro{\v s}kar},
  {Debattista}, {Quinn}, {Stinson}, \& {Wadsley}}]{Roskar-et-al-2008b}
{Ro{\v s}kar}, R., {Debattista}, V.~P., {Quinn}, T.~R., {Stinson}, G.~S., \&
  {Wadsley}, J. 2008{\natexlab{a}}, \apjl, 684, L79

\bibitem[{{Ro{\v s}kar} {et~al.}(2008{\natexlab{b}}){Ro{\v s}kar},
  {Debattista}, {Stinson}, {Quinn}, {Kaufmann}, \&
  {Wadsley}}]{Roskar-et-al-2008a}
{Ro{\v s}kar}, R., {Debattista}, V.~P., {Stinson}, G.~S., {Quinn}, T.~R.,
  {Kaufmann}, T., \& {Wadsley}, J. 2008{\natexlab{b}}, \apjl, 675, L65

\bibitem[{{Salim} {et~al.}(2007){Salim}, {Rich}, {Charlot}, {Brinchmann},
  {Johnson}, {Schiminovich}, {Seibert}, {Mallery}, {Heckman}, {Forster},
  {Friedman}, {Martin}, {Morrissey}, {Neff}, {Small}, {Wyder}, {Bianchi},
  {Donas}, {Lee}, {Madore}, {Milliard}, {Szalay}, {Welsh}, \&
  {Yi}}]{Salim-et-al-2007}
{Salim}, S., {Rich}, R.~M., {Charlot}, S., {Brinchmann}, J., {Johnson}, B.~D.,
  {Schiminovich}, D., {Seibert}, M., {Mallery}, R., {Heckman}, T.~M.,
  {Forster}, K., {Friedman}, P.~G., {Martin}, D.~C., {Morrissey}, P., {Neff},
  S.~G., {Small}, T., {Wyder}, T.~K., {Bianchi}, L., {Donas}, J., {Lee}, Y.-W.,
  {Madore}, B.~F., {Milliard}, B., {Szalay}, A.~S., {Welsh}, B.~Y., \& {Yi},
  S.~K. 2007, \apjs, 173, 267

\bibitem[{{Samland} \& {Gerhard}(2003)}]{Samland-Gerhard-2003}
{Samland}, M. \& {Gerhard}, O.~E. 2003, \aap, 399, 961

\bibitem[{{Schruba} {et~al.}(2011){Schruba}, {Leroy}, {Walter}, {Bigiel},
  {Brinks}, {de Blok}, {Dumas}, {Kramer}, {Rosolowsky}, {Sandstrom},
  {Schuster}, {Usero}, {Weiss}, \& {Wiesemeyer}}]{Schrubaetal2011}
{Schruba}, A., {Leroy}, A.~K., {Walter}, F., {Bigiel}, F., {Brinks}, E., {de
  Blok}, W.~J.~G., {Dumas}, G., {Kramer}, C., {Rosolowsky}, E., {Sandstrom},
  K., {Schuster}, K., {Usero}, A., {Weiss}, A., \& {Wiesemeyer}, H. 2011, \aj,
  142, 37

\bibitem[{{Schuster} {et~al.}(2006){Schuster}, {Marengo}, \&
  {Patten}}]{Schuster-et-al-2006}
{Schuster}, M.~T., {Marengo}, M., \& {Patten}, B.~M. 2006, in Society of
  Photo-Optical Instrumentation Engineers (SPIE) Conference Series, Vol. 6270,
  Society of Photo-Optical Instrumentation Engineers (SPIE) Conference Series

\bibitem[{{Sellwood} \& {Binney}(2002)}]{Sellwood-Binney-2002}
{Sellwood}, J.~A. \& {Binney}, J.~J. 2002, \mnras, 336, 785

\bibitem[{{Swaters} \& {Balcells}(2002)}]{Swaters-Bacells-2002}
{Swaters}, R.~A. \& {Balcells}, M. 2002, \aap, 390, 863

\bibitem[{{Thilker} {et~al.}(2005)}]{Thilker-et-al-2005}
{Thilker}, D.~A. {et~al.} 2005, \apjl, 619, L79

\bibitem[{{Thilker} {et~al.}(2007)}]{Thilker-et-al-2007}
---. 2007, \apjs, 173, 538

\bibitem[{{Trujillo} {et~al.}(2006){Trujillo}, {F{\"o}rster Schreiber},
  {Rudnick}, {Barden}, {Franx}, {Rix}, {Caldwell}, {McIntosh}, {Toft},
  {H{\"a}ussler}, {Zirm}, {van Dokkum}, {Labb{\'e}}, {Moorwood},
  {R{\"o}ttgering}, {van der Wel}, {van der Werf}, \& {van
  Starkenburg}}]{Trujillo-et-al-2006}
{Trujillo}, I., {F{\"o}rster Schreiber}, N.~M., {Rudnick}, G., {Barden}, M.,
  {Franx}, M., {Rix}, H., {Caldwell}, J.~A.~R., {McIntosh}, D.~H., {Toft}, S.,
  {H{\"a}ussler}, B., {Zirm}, A., {van Dokkum}, P.~G., {Labb{\'e}}, I.,
  {Moorwood}, A., {R{\"o}ttgering}, H., {van der Wel}, A., {van der Werf}, P.,
  \& {van Starkenburg}, L. 2006, \apj, 650, 18

\bibitem[{{Werk} {et~al.}(2011){Werk}, {Putman}, {Meurer}, \&
  {Santiago-Figueroa}}]{Werk-et-al-2011}
{Werk}, J.~K., {Putman}, M.~E., {Meurer}, G.~R., \& {Santiago-Figueroa}, N.
  2011, \apj, 735, 71

\bibitem[{{Werk} {et~al.}(2010{\natexlab{a}}){Werk}, {Putman}, {Meurer},
  {Thilker}, {Allen}, {Bland-Hawthorn}, {Kravtsov}, \&
  {Freeman}}]{Werk-et-al-2010-dwarfZ}
{Werk}, J.~K., {Putman}, M.~E., {Meurer}, G.~R., {Thilker}, D.~A., {Allen},
  R.~J., {Bland-Hawthorn}, J., {Kravtsov}, A., \& {Freeman}, K.
  2010{\natexlab{a}}, \apj, 715, 656

\bibitem[{{Werk} {et~al.}(2010{\natexlab{b}}){Werk}, {Putman}, {Meurer},
  {Thilker}, {Allen}, {Bland-Hawthorn}, {Kravtsov}, \&
  {Freeman}}]{Werk-et-al-2010}
---. 2010{\natexlab{b}}, \apj, 715, 656

\bibitem[{{Westra} {et~al.}(2010){Westra}, {Geller}, {Kurtz}, {Fabricant}, \&
  {Dell'Antonio}}]{Westra-et-al-2010}
{Westra}, E., {Geller}, M.~J., {Kurtz}, M.~J., {Fabricant}, D.~G., \&
  {Dell'Antonio}, I. 2010, \pasp, 122, 1258

\bibitem[{{Younger} {et~al.}(2007){Younger}, {Cox}, {Seth}, \&
  {Hernquist}}]{Younger-et-al-2007}
{Younger}, J.~D., {Cox}, T.~J., {Seth}, A.~C., \& {Hernquist}, L. 2007, \apj,
  670, 269

\bibitem[{{Zaritsky} \& {Christlein}(2007)}]{Zaritsky-Christlein-2007}
{Zaritsky}, D. \& {Christlein}, D. 2007, \aj, 134, 135

\bibitem[{{Zibetti} \& {Groves}(2011)}]{Zibetti-Groves-2011}
{Zibetti}, S. \& {Groves}, B. 2011, \mnras, 417, 812

\end{thebibliography}

\end{document}